# The Impact of Sex Education on Sexual Activity, Pregnancy, and Abortion

Nima Khodakarami

Health Policy and Management

Texas A&M School of Public Health

**Abstract**

The purpose of this study is to find a relation between sex education and abortion in the United States. Accordingly, multivariate logistic regression is employed to study the relation between abortion and frequency of sex, pre-marriage sex, and pregnancy by rape. The finding shows the odds of abortion among those who have had premarital sex, more frequent sex before marriage, and been the victim of rape is higher than those who have not experienced any of these incidents. The output identified with one unit increase in pre-marriage sex the log-odds of abortion increases by 0.47. Similarly, it shows by one unit increase in the frequency of sex, the log-odds of abortion increases by 0.39. Also, for every additional pregnancy by rape, there is an expectation of a 3.17 increase in the log-odds of abortion. The findings of this study also suggests abortion is associated with sex education. Despite previous findings, this study shows the factors of age, having children, and social standing is not considered a burden to parents and thereby do not have a causal relation to abortion.

## Introduction

In the countries where abortion is highly restricted by law, unsafe abortion has been endangered life of many women. On the other hand in the liberal countries, safe abortion has neither been easily accessible (Grimes et al., 2006) nor harmless. No matter of what is the type of abortion, there are public health issues involved e.g. feeling doubtful, sad, confused, or guilty about the abortion (Skowronski, 1977) as well as the risk of mental issue (Fergusson, John Horwood, & Ridder, 2006) and breast cancer among the young women (HOWE, SENIE, BZDUCH, & HERZFELD, 1989).To overcome the issue of unintended pregnancy, the U.S. government have funded sex education in public schools of the large cities aiming for a reduction in harmful aspects of unintended pregnancy. The program has started decades ago and was offered either as an individual course or as part of the curriculum and covered more than three-quarter of teenagers in the large U.S. cities (Marsiglio, 1986). As of 1984, more than fifty percent of young men and women who have been at their 20[th] had received sex education by age of 19. The support for the program has received from both teachers and parents who have been in favor of sex education however, this course of study has always had its controversial issues along with opponents and proponents. The opponents of sex education have claimed it soars the possibility of sexual activity and pregnancy among teenagers. This argument which is backed by a recent study that occurred within a time period of 2008–2011 and showed U.S. teenagers of 15-19 year old has had the highest pregnancy rate comparing to other 21 developed countries (Sedgh, Finer, Bankole, Eilers, & Singh, 2015) has been denied by the proponents of sex education. The supporters have argued sex education reduces the rate of pregnancy by through delivery of contraceptive knowledge and impregnation practices (Marsiglio, 1986).

To better identify the points of controversy, several empirical studies are reviewed below. These empirical researches have been seeking weather sex education have had a prohibitive or permissive role in teenagers sexual activity and pregnancy claim, which reported their findings as below.

One study found a little relationship between premarital sexual activity and sex education in the adolescent group (Marsiglio, 1986). Another national level study showed sex education decreases the level of sexual activity among teenagers of 15-16 years old (Marsiglio, 1986). Zelnik and Kim (1979) found a similar outcome among 15-19 years old teenagers of metropolitan areas. They further showed their people of study are less likely to experience a premarital pregnancy (Zelnik, 1979).  Marsiglio (1986) analyzed a longitudinal Survey of Work Experience of Youth (NLSY) in which 6,015 women and 6,054 men aged 19-27 have been responded to the survey interviews. They concluded sex education increases sexual activity among 15-16 years old teenagers however, it has an incremental effect on the likelihood of contraceptive use (Marsiglio, 1986). Shah & Zelnik (1981) analyzed the premarital sexual behavior of 15-19 years old women. They found women that were influenced by their peers had a high level of premarital pregnancy (Shah, 1981).

Where pregnancy rate has been impacted by the course of sex education, nine reasons were cited to be common for terminating pregnancy including: schooling and fear of expulsion, low income and unaffordable child care, social condemnation where premarital pregnancy is immoral, having no stable relationship, failed contraceptive use, rape incidents, having a child, dislike toward father of baby, and forced to abortion (Olukoya, Kaya, Ferguson, & AbouZahr, 2001). It is also asserted that the major reasons for abortion are financial difficulties and lack of a partner (Lawrence et al., 2005). Lawrence, Frohwirth, Dauphinee, Singh, & Moore (2005) have also shown the decision to have an abortion is typically motivated by the cost of raising children, responsibilities to children, partner issues and unreadiness to be a parent (Lawrence et al., 2005). Stanger-Hall et al. (2011) examined data on pregnancy, birth and abortion rates of female teens between 15 and 19 years of age for 48 states (all U.S. states except North Dakota and Wyoming) through 2005 (Stanger Hall, Hall, & Vitzthum, 2011).

As I discussed above sex education proved to have a positive impact on unwanted pregnancy. However, "Between 2001 and 2008, intended pregnancies decreased and unintended pregnancies increased" (Finer & Zolna, 2014) that implies, "the appropriate type of sex education that should be taught in U.S. public schools continues to be a major topic of debate, which is motivated by the high teen pregnancy and birth rates in the U.S., compared to other developed countries" (Stanger Hall et al., 2011). That justifies studies being more focused on the efficiency of sex education course and attempt to discover its correlation with sexual activity and pregnancy. But would the sex education impact abortion remained unclear? Considering the U.S. has ranked 15 among the 21 developed countries in terms of abortion, while it is number 1 among the same countries in terms of teenager's pregnancy (Sedgh et al., 2015), I found it interesting to seek for the possible impact of sex education on abortion.

This study, to my knowledge, will be the first one of its type that attempts to examine the relationship between sex education and rate of abortion. This study is structured as follows. First, I develop a hypothesis based on the current literature. Then, I will argue my method and data collection. Finally, I will estimate a multivariate logit model and report my result.

## Hypothesis Development

Prior empirical studies have analyzed the impact of sex education on pregnancy and sexual activity. While there has been consent on the increasing impact of education on sexual activity, the argument will be built on the correlation between sex education and sexual activity. Firstly, I use sexual activity as a proxy for sex education. By increase in sexual activity the likelihood of unwanted pregnancy increases (Marsiglio, 1986). While a great percentage of unwanted pregnancy end in abortion (Finer & Zolna, 2014) there is an expectation that an increase in sexual activity leads to more abortion. However, a huge gap between pregnancy rate and abortion in the U.S. ranking (Sedgh et al., 2015) implies there is no relation between abortion and sex education. Thereupon, I will be examining the existence of the dotted line from sex education to abortion in figure 1.

*Hypothesis:* Exposure to formal sex education has no impact on the abortion.

**Figure 1**: Causal Diagram of Abortion

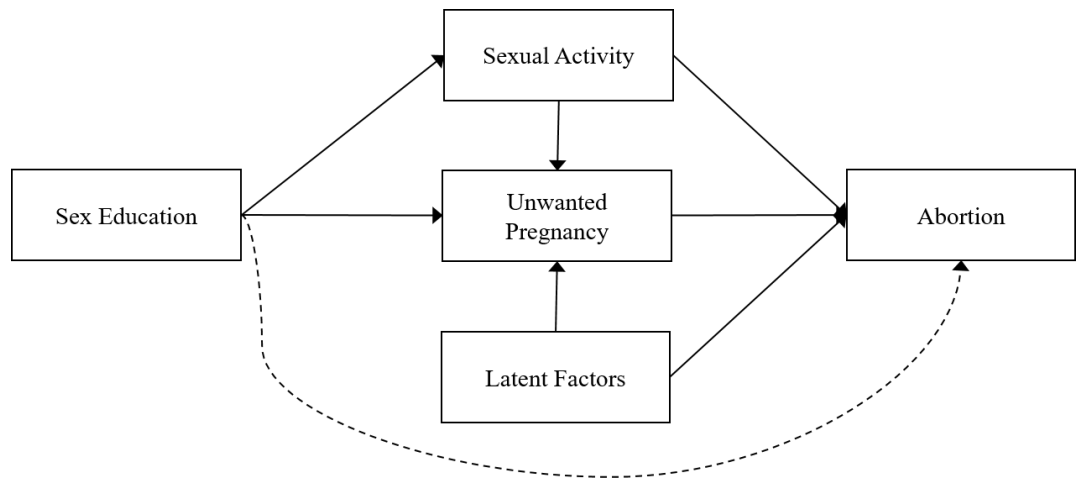

## Methodology

This study aims to measure the correlation between sex education and abortion. For this objective, I supplemented my study with data coming from the contemporary American society available as part of the National Data Program for Social Science. The data was collected through the General Social Survey conducted by NORC at the University of Chicago. Demographic, behavioral and attitudinal responses have been recorded in the survey. Access to the corresponding data is allowed at the GSS Data Explorer, stored in SPSS and STATA format for download. Extra sources for accessing the data are SDA, Roper Center and ICPSR. The individual year data sets in the GSS are only in the cross-sectional format and covers the period of 1972 to 2016 (Smith, Davern, Freese, and Hout, 1972:2016).

I will use the logit regression approach to assess the relationship between sex education and abortion. I will use pre-marriage sex and sex frequency as proxies of increase in sexual activity. To stay consistent with the previous proposal, I use data of 2012. This set of data contains 1974 observation and 859 variables. I pick following factors as the right-hand side variables: highest years of school completed by respondent, total family income, the financial burden of children on parents, relationship status and cohabitation, birth control to teenagers 14-16, pregnancy as a result of rape, sex before marriage, and frequency of sex. The dependent variable is "abortion if women want for any reason".

## Design and Procedures

The data was downloaded in a STATA format and was used for further analysis. The following measures were employed within a regression model.

### *Independent Variables*

Two independent variables are chosen for the objective of the study. One is inferred from the prior literature, where studies agree on the effect of sex education. As they consent on the positive effect of sex education on sexual activity, the following independent variable is used as a proxy for sex education:

*Sex frequency.* The frequency of sex has commonly used as a good measure of sexual activity in prior studies. For instance, Kirby et.al. (2007) seeks frequency of sex in a specific period of time prior to their survey. Their sex frequency question asks "whether or not respondents had sex at all during that period of time" (Kirby, Laris, & Rolleri, 2007).

Despite the consent on the impact of sex education on sex frequency, prior studies have not reached to an agreement about the effect of sex education on premarital sex.

*Pre-marriage sex.* Increase or initiation of premarital sex has been a controversial outcome for sex education. For example, Dawson (1986) did not find any consistent relationship between sex education and the subsequent initiation of intercourse (Dawson, 1986). However, another study argues sex education influence premarital sex (Marsiglio, 1986).

### Control variables

In a regression model, I controlled for a number of variables that might impact abortion. These control variables are the most common causes of unwanted pregnancy. For example, according to Finer and Zola (2014) use of contraceptive methods within unequal socioeconomic communities may induce lower unwanted pregnancy (Finer & Zolna, 2014). Also, studies showed women with lowers years of schooling had the highest unwanted abortion. The rate of unwanted pregnancy has also increased among poor and low income women, cohabiters and formerly married women (Finer & Zolna, 2011). To seize the effect of all discussed variables the control variables are:

*Birth control.*  Comparing sexual activity with data on contraceptive use, the latter variable could be a better explanatory variable for reasoning the level of pregnancy and abortion among adolescents (Darroch, 2001). Because of that, I added this variable to the set of control variables.

*Education.* I employed one variable of education to measure the level of schooling of the respondents. Since I want to find linear relationships, I justified the right-skewed education by grouping the education level into four group of kindergarten to middle school (0-8th years of schooling), high school (9th to 12th years of schooling), undergraduate level (13th-16th years of schooling), and graduate level (17th -20th years of schooling).

*Income.* There have been several variables reflecting the income of respondents and his/her family. The variety of income variables contained information like high income, total family income, respondent's income at her 16 years old, etcetera. For this study I used total family income that was categorized in 25 ranges of earning. Again, to remove the skewness from data, I grouped the revenue in seven set. To do a meaningful grouping, I checked the percent distribution of total income for the households in the United States (*See* https://www.census.gov for more detail).

*Relationship status.* I followed prior studies that noted relationship status induces abortion. According to previous studies, by an increas of cohabitation unwanted pregnancy has increased. A large shift from intended to unwanted pregnancy has then ended in higher abortion (Finer & Zolna, 2011; Finer & Zolna, 2014). Additionally, I included three other variables, which are common motives for abortion to the control variables (Olukoya et al., 2001):

*Rape victim*. Because it is necessary to consider the reason for abortion, rape is another variable for this study. Typically, rape is known as a second thread to abortion (Subrahmanyam, Greenfield, & Tynes, 2004). As reported in this set of data, pregnancy as a result of rape recorded as a Yes/No answer.

*Financial burden*. In line with income, I used a measure that shows how respondents measure children as a financial burden on parents in a Likert scale. To create reverse-score items, I transformed this data to 0-4 increasing scale.

*Social standing*. To measure the social condemnation factor, I used a variable reflecting views on social standing in society as a consequence of having children.  This variable has also display responses in a Likert five-point level.

*Age*. Coleman (2006) showed about one-quarter of abortion in the United States occur on women under age 20. That make age an interesting variable for inclusion in the regression model.

### Logit regression model for GSS-2012 sample

To estimate the relationship between the binary dependent variable and the independent variables, I used logistic (LOGIT) regression analysis. The result of regression is reported in below. I coded the dependent variable as a "1' for those who replied yes to the survey question of "abortion if women want for any reason" and "0" for the no responses. Then the regression model would be:

$\log(\frac{p}{1-p}) = \beta_{intercept} + \beta_{education}x_1 + \beta_{income}x_2 + \beta_{finanbrdn}x_3 + \beta_{relationstatus}x_4 + \beta_{brthctrl}x_5 + \beta_{rape}x_6 + \beta_{socialstat}x_7 + \beta_{age}x_8 + \beta_{premarsex}x_9 + \beta_{sexfreq}x_{10} + \mu$

**Results**

The postulated hypothesis predicted that there is no relation between sex education and abortion. I conducted a logit regression model to test this hypothesis. Total of 1,248 interviewees responded to the abortion question. However, when data has been used along with other variables the number of observation has dropped to 247, 244, and 165. That could be a cause of missing data in each set of responses of the independent variables or the low number of respondents.

Sum of 554 have replied "yes" to the abortion and 694 said "no" to the abortion. Looking at the variables explaining unwanted pregnancy, the sum of 1,230 pregnant women has participated in the survey. The percentage of pregnancy as a result of rape has been 76.5 percent equals to a total of 941, compared to 23.5 percent, not the victims. The significant variable rape explains for every additional pregnancy by rape we expect a 3.17 increase in the log-odds of abortion. The output shows the likelihood of abortion as a result of rape is pretty steady with and without sex education. Similarly, the birth control variable is significant in the first and second regression. This variable explains for every unit increase of intention to use the contraceptive method the log-odds of abortion increases by 0.37. When we add the variable premarital sex the odd ratio of abortion drops. However, this variable shows no more significant influence when we add the variable sex frequency. Other control variables have not had a significant coefficient that explains there is no relationship between the log-odds of abortion and those variables (*See* Table 1).

Despite the expectation, premarital sex has been significant. The output shows with one unit change in the belief that sex before marriage is always wrong to almost always wrong, sometimes wrong, and not wrong at all, the log-odds of abortion increases to 0.47. Similarly, for the frequency of sex during the past year, by one unit increase in the frequency of intercourse, we expect 0.39 escalation in the log-odds of abortion. (The units of this variable are not having sex at all in the past year, once or twice over the last year, once a month, 2-3 times a month, weekly, 2-3 times per week and more than 4 times per week.). With the recent model the expectation for the log-odds of abortion increases by 0.13% (from 0.47 to 0.53)

**Table 1**: Logistic Regression Output

| VARIABLES | (1) abortion | (2) abortion | (3) abortion |
|---|---|---|---|
| education | 0.30 | 0.29 | 0.10 |
| | (0.244) | (0.258) | (0.299) |
| income | 0.11 | 0.08 | 0.15 |
| | (0.108) | (0.106) | (0.125) |
| financial burden | 0.15 | 0.10 | 0.11 |
| | (0.153) | (0.162) | (0.207) |
| relationship status | 0.16 | 0.13 | 0.19 |
| | (0.142) | (0.144) | (0.205) |
| birth ctrl | 0.37** | 0.32* | 0.25 |
| | (0.174) | (0.178) | (0.219) |
| rape victim | 3.62*** | 3.58*** | 3.17*** |
| | (0.711) | (0.843) | (0.861) |
| social standing | 0.01 | -0.00 | -0.11 |
| | (0.187) | (0.195) | (0.241) |
| age | 0.00 | 0.01 | 0.03 |
| | (0.009) | (0.010) | (0.018) |
| premarital sex | | 0.47** | 0.53** |
| | | (0.191) | (0.256) |
| sex frequency | | | 0.39** |
| | | | (0.158) |
| Constant | -5.52*** | -6.56*** | -7.77*** |
| | (1.118) | (1.491) | (1.710) |
| | | | |
| Observations | 247 | 244 | 165 |

Robust standard errors in parentheses

*** $p<0.01$, ** $p<0.05$, * $p<0.10$

**Discussion**

Results indicate that sex frequency and premarital sex are the fuels of abortion. The analysis of data for 2012 showed the variables of age, children as a social burden to parents and social standing have not had a causal relation with abortion. Although the bivariate regression of the noted variables shows a significant relation the multivariate result denotes no or little relation between abortion and those variables. However, corresponding to the findings of prior studies I found abortion as a significant factor in abortion. This also goes along with the factor of hatred or dislike toward the father of baby [10]. Although, due to a limitation of data, access to this variable was restricted. So, the effect of dislike toward father could be inherited in the rape variable. Additionally, the finding implies the use of birth control methods in cases the sex frequency is high has no significant relation with abortion. It could imply a correlation between variables, but pre regression collinearity test showed no collinearity among the independent variables.

The premarital sex analysis along with sex frequency analysis revealed a new relationship between sex education and abortion. The output of significant variables rejects our null hypothesis that there is no relation between sex education and abortion. But, it draws a new line of cautious that how resourceful and efficient is the sex education to public health. With this finding, the alternative hypothesis of sex education impacts abortion has been justified.

This study has also had some caveats. Firstly, the above study examined women of various ages. While the impact of sex education varies between adolescents and teenagers, the output could not be a good indicator of the impact of sex education on abortion. Therefore a more detailed study containing specific subsample of teenagers and adolescents is suggested for future study.

While the finding of the study casts doubt on the efficiency of sex education, it should be noted that data limitation may affect the validity and usefulness of this finding. Even more, the probable relation between sex education and abortion could be a result of correlation, not causation. Hence, a test for the indigeneity of this issue is required.

In conclusion, some of the variables defined as a causal factor to abortion or pregnancy showed no relation with the abortion. This finding suggests a new study incorporating al the variables to reduce the possible issue of the omitted variable in the regression.

**References**

Darroch, J. E. (2001). Differences in teenage pregnancy rates among five developed countries: The roles of sexual activity and contraceptive use. *Family Planning Perspectives,* , 244. doi:10.2307/3030191

Dawson, D. A. (1986). The effects of sex education on adolescent behavior. *Family Planning Perspectives,* , 162. doi:10.2307/2135325

Fergusson, D., John Horwood, L., & Ridder, E. (2006). Abortion in young women and subsequent mental health. *Journal of Child Psychology and Psychiatry, 47*(1), 16-24. doi:10.1111/j.1469-7610.2005.01538.x

Finer, L., & Zolna, M. (2011). Unintended pregnancy in the united states: Incidence and disparities, 2006. *Contraception, 84*(5), 478-485. doi:10.1016/j.contraception.2011.07.013

Finer, L., & Zolna, M. (2014). Shifts in intended and unintended pregnancies in the united states, 2001:2008. *American Journal of Public Health, 104*(S1), S48. doi:10.2105/AJPH.2013.301416

Grimes, D., Benson, J., Singh, S., Romero, M., Ganatra, B., Okonofua, F., & Shah, I. (2006). Unsafe abortion: The preventable pandemic. *The Lancet, 368*(9550), 1908-1919. doi:10.1016/S0140-6736(06)69481-6

HOWE, H., SENIE, R., BZDUCH, H., & HERZFELD, P. (1989). Early abortion and breast cancer risk among women under age 40. *International Journal of Epidemiology, 18*(2), 300-304. doi:10.1093/ije/18.2.300

Kirby, D., Laris, B. A., & Rolleri, L. (2007). Sex and HIV education programs: Their impact on sexual behaviors of young people throughout the world. *Journal of Adolescent Health, 40*(3), 206-217. doi:10.1016/j.jadohealth.2006.11.143

Lawrence, B., Finer, L., Frohwirth, L., Dauphinee, A., Singh, S., & Moore, A. (2005). Reasons U.S. women have abortions: Quantitative and qualitative perspectives. *Perspectives on Sexual and Reproductive Health, 37*(3), 110-118. doi:10.1111/j.1931-2393.2005.tb00045.x

Marsiglio, W. (1986). The impact of sex education on sexual activity, contraceptive use and premarital pregnancy among american teenagers. *Family Planning Perspectives, , 151.* doi:10.2307/2135324

Olukoya, A. A., Kaya, A., Ferguson, B. J., & AbouZahr, C. (2001). Unsafe abortion in adolescents. *International Journal of Gynecology and Obstetrics, 75*(2), 137-147. doi:10.1016/S0020-7292(01)00370-8

Sedgh, G., Finer, L., Bankole, A., Eilers, M., & Singh, S. (2015). Adolescent pregnancy, birth, and abortion rates across countries: Levels and recent trends. *Journal of Adolescent Health, 56*(2), 223-230. doi:10.1016/j.jadohealth.2014.09.007

Shah, F. (1981). Parent and peer influence on sexual behavior, contraceptive use, and pregnancy experience of young women. *Journal of Marriage and Family, , 339.* doi:10.2307/351385

Skowronski, M. (1977). *Abortion and alternatives*. Millbrae, Calif: Femmes.

Smith, Tom W., Davern, Michael, Freese, Jeremy, and Hout, Michael, General Social Surveys, 1972-2016 [machine-readable data file] /Principal Investigator, Smith, Tom W.; Co-Principal Investigators, Peter V. Marsden and Michael Hout; Sponsored by National Science Foundation. --NORC ed.-- Chicago: NORC, 2018: NORC at the University of Chicago [producer and distributor]. Data accessed from the GSS Data Explorer website at gssdataexplorer.norc.org.

Stanger Hall, K., Hall, D., & Vitzthum, V. (2011). Abstinence-only education and teen pregnancy rates: Why we need comprehensive sex education in the U.S. *PLoS One, 6*(10), e24658. doi:10.1371/journal.pone.0024658

Subrahmanyam, K., Greenfield, P., & Tynes, B. (2004). Constructing sexuality and identity in an online teen chat room. *Journal of Applied Developmental Psychology, 25*(6), 651-666. doi:10.1016/j.appdev.2004.09.007

Zelnik, M. (1979). Sex education and knowledge of pregnancy risk among US teenage women. *Family Planning Perspectives, 11*(6), 355. doi:10.2307/2134219